\documentclass[prl,twocolumn,superscriptaddress,nobibnotes,letterpaper]{revtex4-1}

\usepackage[pdftex]{graphicx}
\usepackage[pdftex]{epsfig}
\usepackage{amsmath}
\usepackage{amssymb}
\usepackage{amsfonts}
\usepackage{color}
\usepackage{wrapfig}
\usepackage{eucal}
\usepackage{hhline}
\usepackage{threeparttable}
\usepackage{supertabular}
\usepackage{multirow}
\usepackage{tabularx}
\usepackage{upgreek}

\usepackage{soul}

\usepackage{array} % make column bigger (vertical)
\usepackage[colorlinks=true, allcolors=blue]{hyperref} %hyperrefs

\newcommand{\und}[1]{_\textrm{#1}}

\definecolor{cgreen}{rgb}{.1,.6,.1}
\definecolor{co}{rgb}{.1,.6,.6}
\definecolor{orange}{rgb}{.9,.4,.0}

\newcolumntype{C}[1]{>{\centering\arraybackslash}p{#1}}

\begin{document}
\pagenumbering{arabic}

\title{Feedback cooling of a room temperature mechanical oscillator\\ close to its motional groundstate}\thanks{This work was published in \href{https://doi.org/10.1103/PhysRevLett.123.223602}{Phys.\ Rev.\ Lett.\ \textbf{123}, 223602} (2019). The source data for the figures is available at \href{https://doi.org/10.5281/zenodo.3906185}{10.5281/zenodo.3906185}.}

\author{Jingkun Guo}
\affiliation{Kavli Institute of Nanoscience, Department of Quantum Nanoscience, Delft University of Technology, 2628CJ Delft, The Netherlands}

\author{Richard Norte}
\email{r.a.norte@tudelft.nl}
\affiliation{Kavli Institute of Nanoscience, Department of Quantum Nanoscience, Delft University of Technology, 2628CJ Delft, The Netherlands}
\affiliation{Department of Precision and Microsystems Engineering, Delft University of Technology, Mekelweg 2, 2628CD Delft, The Netherlands}

\author{Simon Gr\"oblacher}
\email{s.groeblacher@tudelft.nl}
\affiliation{Kavli Institute of Nanoscience, Department of Quantum Nanoscience, Delft University of Technology, 2628CJ Delft, The Netherlands}

%\date{\today}

\begin{abstract}
Preparing mechanical systems in their lowest possible entropy state, the quantum groundstate, starting from a room temperature environment is a key challenge in quantum optomechanics. This would not only enable creating quantum states of truly macroscopic systems, but at the same time also lay the groundwork for a new generation of quantum limited mechanical sensors in ambient environments. Laser cooling of optomechanical devices using the radiation pressure force combined with cryogenic pre-cooling has been successful at demonstrating groundstate preparation of various devices, while a similar demonstration starting from a room temperature environment remains an outstanding goal. Here we combine integrated nanophotonics with phononic bandgap engineering to simultaneously overcome prior limitations in the isolation from the surrounding environment, the achievable mechanical frequencies, as well as limited optomechanical coupling strength, demonstrating a single-photon cooperativity of 200. This new microchip technology allows us to feedback cool a mechanical resonator to around 1~mK, near its motional groundstate, from room temperature. Our experiment marks a major step towards accessible, widespread quantum technologies with mechanical resonators.
\end{abstract}

\maketitle

The last decade has seen immense progress on observing quantum effects with microfabricated mechanical oscillators~\cite{OConnell2010,Palomaki2013,Hong2017,Riedinger2018,Ockeloen-Korppi2018,Marinkovic2018}. This is not only of significant interest for understanding the fundamental aspects of quantum physics in macroscopic objects, but also for the potential of using mechanical systems for quantum information processing tasks and as novel quantum sensors~\cite{Norte2018}. Excess classical (i.e.\ thermal) noise typically obscures the quantum features of these devices, thus limiting their usefulness and practical adoption for quantum limited sensing. Groundstate cooling can alleviate this problem, but so far has only been possible by pre-cooling the devices using cryogenic methods~\cite{Chan2011,Teufel2011b,Underwood2015,Peterson2016,Rossi2018}. The main limitations preventing to reach this regime from room temperature include insufficient isolation from the surrounding environment and too low mechanical frequencies, which can be formulated into the condition of the product of the mechanical frequency and its quality factor $f\und{m}\cdot Q\und{m}\geq6\times10^{12}$~\cite{Marquardt2007}. In addition, the optomechanical coupling rate $g_0$ also plays a dominant role in the ability to efficiently laser cool the motion of a resonator. There are several approaches focusing on overcoming these limitations. In particular, experiments featuring optically levitated nanospheres have come to within a few thermal phonons of the mechanical groundstate recently~\cite{Delic2019,Tebbenjohanns2019a,Conangla2019,Tebbenjohanns2019b}. While the absence of any physical attachment to the environment allows trapped nanospheres to exhibit extremely large quality factors, they require UHV systems and complex stabilization mechanisms for their optical traps, making them impractical as sensors and for other applications. Chip-based mechanical oscillators have recently also been shown to feature competitively large mechanical quality factors at room temperature with $Q\und{m}\gtrsim10^8$, most prominently in high-stress silicon nitride membranes~\cite{Norte2016,Reinhardt2016,Tsaturyan2017}. Here, similar limitations as with levitated nanospheres, such as mirror noise~\cite{Liu2000,Harry2006}, exist, as well as the requirement to use bulky setups for optical readout.

\begin{figure*}[t!]
	\includegraphics[width=2\columnwidth]{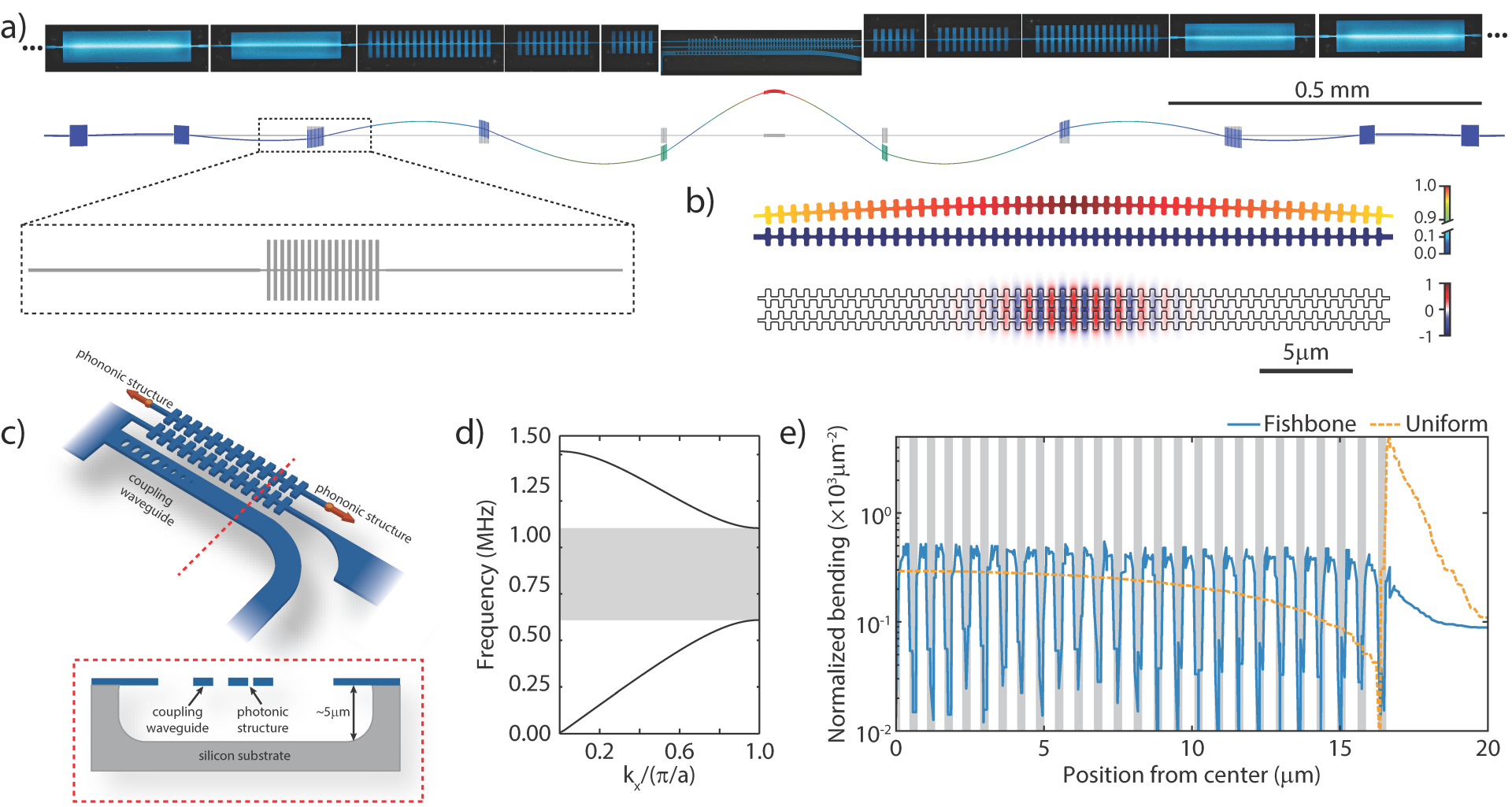}
	\caption{(a) Shown is a stitched microscope image of the fabricated structure (top) and the corresponding mechanical simulation of the long beam (bottom). The zoom-in shows a fishbone structure designed to reduce bending losses. (b) Mechanical (top) and photonic (bottom) simulation of the center part containing the photonic crystal. A short second beam forming the other half of the photonic crystal cavity is fixed close to the mechanical beam. The short structure does not feature any mechanical motion around the defect-mode. The two structures form an optical cavity with the light strongly confined in the gap between the two beams. The mechanial motion changes the gap size, shifting the optical resonance frequency and hence giving rise to the optomechanical interaction. (c) Sketch of the central photonic crystal and coupling waveguide. Clearly visible is the short second beam in the center. A cross-section is shown in the box on the bottom. (d) Band diagram for the in-plane mode of the phononic fishbone structure. We apply a periodic boundary condition in the $x$-direction, with $k_x$ being the wavevector. The blocks in the long beam form a bandgap between 0.6 and 1.1~MHz. (e) Bending ($\partial^2 v/\partial x^2$) normalized to the displacement $v$ within the photonic fishbone structure. In the center, the magnitude of the bending alternates between the thin (large bending, white) and thick (small bending, gray) parts of the structure. As the mechanical losses are proportional to the cube of the width, the fishbone devices exhibit significantly higher mechanical $Q\und{m}$ than a uniform beam of equal width (orange), typically used for photonic crystal (PhC) zipper cavities~\cite{Krause2015}.}
	\label{fig:1}
\end{figure*}

In this work, we develop a new type of fully integrated optomechanical structure that allows us to significantly increase the mechanical quality factor of a high-frequency in-plane mode, while also allowing to realize a coupled opto-mechanical cavity used for on-chip optical read-out of the motion. We measure a $f\und{m}\cdot Q\und{m}\approx2.6\times10^{13}$, approaching the performance of the best out-of-plane mechanical resonators~\cite{Ghadimi2018,Rossi2018}, combined with an optomechanical coupling of $G\und{om}/2\pi=21.6\pm0.2$~GHz/nm, enabling us to cool the mechanical mode from room temperature to 1.2~mK. This corresponds to a thermal mode occupation of less than 27 phonons, a reduction by more than 5 orders of magnitude in the effective temperature. Our novel design applies previous discoveries on the dominant role of bending losses~\cite{Unterreithmeier2010, Ghadimi2018} and results in a device that resembles a fishbone-like photonic and phononic structure (cf.\ Figure~\ref{fig:1}a).

Significant progress has been made over the last years in understanding and mitigating the losses in integrated (opto-) mechanical structures, resulting in experimental demonstrations of ultra-high $Q\und{m}$ devices. In particular, bending loss has been shown to be one of the dominant limiting  mechanisms for mechanical quality factors in 1D high-stress silicon nitride structures~\cite{Unterreithmeier2010}. Various approaches in strain~\cite{Norte2016,Reinhardt2016} and mode shape engineering~\cite{Ghadimi2018} have recently succeeded in achieving ultrahigh-$Q\und{m}$ mechanical resonators. By using adiabatically chirped phononic crystals for example~\cite{Ghadimi2018}, the mechanical mode is localized in the center of the beam and the bending can significantly be reduced, leading to increases in $Q\und{m}$. While this concept works very well for the out-of-plane motion, it is much more challenging for an in-plane mode~\cite{GhadimiPhD}. This is due to the loss $\Delta U$ being proportional to the cube of the thickness in the motional direction~\cite{Ghadimi2018}, which for the in-plane mode is equivalent to the width $w$ of the structure $\Delta U\propto w^3(\partial^2 v/\partial x^2$), with $v$ being the displacement. In practice, there are several parts of an optomechanical structure that require a certain minimum width, such as the phononic crystal itself, which is partly comprised of wide blocks of material. The bending of these very wide blocks results in large mechanical dissipations. Furthermore, in order to form a good optical cavity, the photonic crystal at the center of the structure also requires a minimum width, which is comparable to the optical wavelength~\cite{Krause2015}. Both factors largely reduce the attainable mechanical quality factor. With our new fishbone design we minimize $w$ in the parts with maximal bending, allowing us to significantly reduce $\Delta U$, and hence significantly increase the mechanical quality factor of the mechanical in-plane modes.

Our structure is fabricated from a 350~nm thick high-stress (1.3 GPa) silicon nitride layer deposited on a silicon handle wafer. As shown in Figure~\ref{fig:1}a, it is based on the differential motion of two strings, where one of them is significantly longer (2.6~mm) than the other (115~$\mu$m). The longer string of this zipper structure is connected to the chip through a phononic crystal, with a bandgap for the in-plane mode between 610~kHz and 1.10~MHz (see Figure~\ref{fig:1}d). This design forms a defect in the center, introducing confined modes with frequencies within the bandgap, significantly reducing the losses of these modes. As the amplitude of the modes of interest is largest in the center of the structure, we reduce its bending by introducing an adiabatic transition of the unit cells of the phononic crystals. This results in a weaker confinement and smaller bending close to the center~\cite{Ghadimi2018}. As mentioned above, this does however not immediately result in a high quality factor of the in-plane modes, as the width of the structure close to the bending areas becomes important. We therefore design the overall device as a string with a width of only 165~nm, limited by our fabrication process. When adding the phononic crystal we avoid wide and rigid regions in the design, segmenting the blocks closest to the center into a fishbone-like structure.

A similar approach is also taken for the central photonic crystal used to read-out the mechanical motion. Instead of a traditional photonic structure with holes in a wavegeuide~\cite{Krause2015}, we achieve an alternating index contrast through a fishbone design. The wider parts are roughly 1~$\mu$m in width, while the narrow ones between the teeth have a width of only 165~nm. This geometry localizes the bending to the narrow parts (cf.\ Figure~\ref{fig:1}e), significantly reducing the overall bending losses. For comparison, we observe a typical enhancement of $f\und{m}\cdot Q\und{m}$ by more than a factor of 5 between devices with and without the photonic fishbone structure.

The optical cavity is formed between the long and short strings and the optical field is confined within the gap formed by the fishbones and designed to operate at a wavelength of around $\lambda=1550$~nm. Due to the strong confinement, the resonance frequency of the cavity is very sensitive to the gap size. In the simulation, shown in Figure~\ref{fig:1}b, we obtain an optomechanical coupling strength $G\und{om}/2\pi = \frac{\partial \omega\und{c}}{\partial x} = 23.0$~GHz/nm, with $\omega\und{c}$ the cavity frequency, for a typical 200~nm gap. Combining $G\und{om}$ with the localized mechanical mode of interest, which has small effective mass $m\und{eff}=7.36\times10^{-14}$~kg, we obtain a single photon coupling rate $g_0/2\pi=252$~kHz. The resulting optimized structure features a 16.5~$\mu$m long photonic crystal cavity, while the overall structure has a length of 2.6~mm.

We design the mechanical defect mode at a frequency $f\und{m}=\omega\und{m}/2\pi=950$~kHz. A ringdown measurement of this mode in $5\times10^{-6}$~mbar vacuum, shows a quality factor of $2.73\times10^7$ (cf.\ SI), yielding $f\und{m}\cdot Q\und{m}=2.59\times10^{13}$. The total optical resonance's ($\lambda=1549.9$~nm) linewidth is measured to be $\kappa/2\pi=33.0$~GHz, and the coupling rate to an adjacent optical waveguide $\kappa\und{e}/2\pi=31.4$~GHz. The strongly over-coupled cavity ensures that most of the light in the cavity is reflected back into the waveguide, which is necessary to achieve high detection efficiency. To further characterize the device, we measure the optical spring effect (see SI for details), allowing us to experimentally determine a single photon coupling rate of $g_0/2\pi=237\pm2$~kHz, corresponding to an optomechanical coupling of $G\und{om}/2\pi=21.6\pm0.2$~GHz/nm, in good agreement with simulations. We determine the single photon cooperativity of our device $C_0=\frac{4g_0^2}{\kappa\Gamma\und{m}}\approx200$, which represents the relative strength of the single photon interactions against any loss channels, a key characteristic of the system~\cite{Krause2015,Wilson2015}.

\begin{figure}[t!]
	\includegraphics[width=1.\columnwidth]{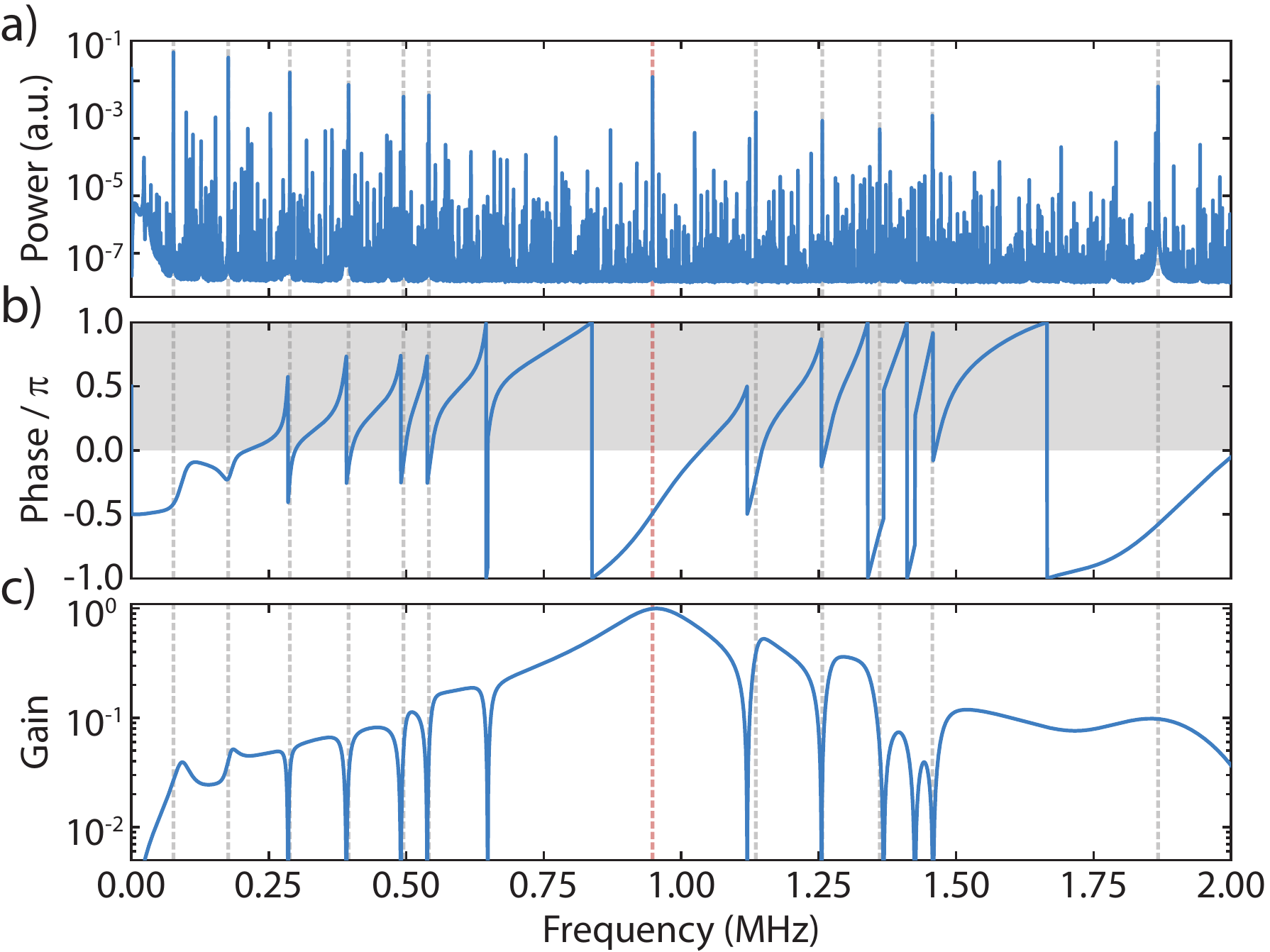}
	\caption{(a) Mechanical spectrum. The red dashed line marks the high-$Q\und{m}$ defect mode. Gray lines indicate additional mechanical modes that strongly couple to the optical cavity, while most other spurious peaks arise from the mixing of these modes in the detection itself. (b) Phase response of the feedback control. The circuit has a phase of $-0.5\pi$ at the resonance of the defect mode, with the gray area indicating the unstable region due to heating. (c) Gain of the feedback control. We implement several filter functions with large bandwidth, allowing us to suppress and partly cool other modes that are excited in order to stabilize the system.}
	\label{fig:3}
\end{figure}

\begin{figure}[t!]
	\includegraphics[width=1.\columnwidth]{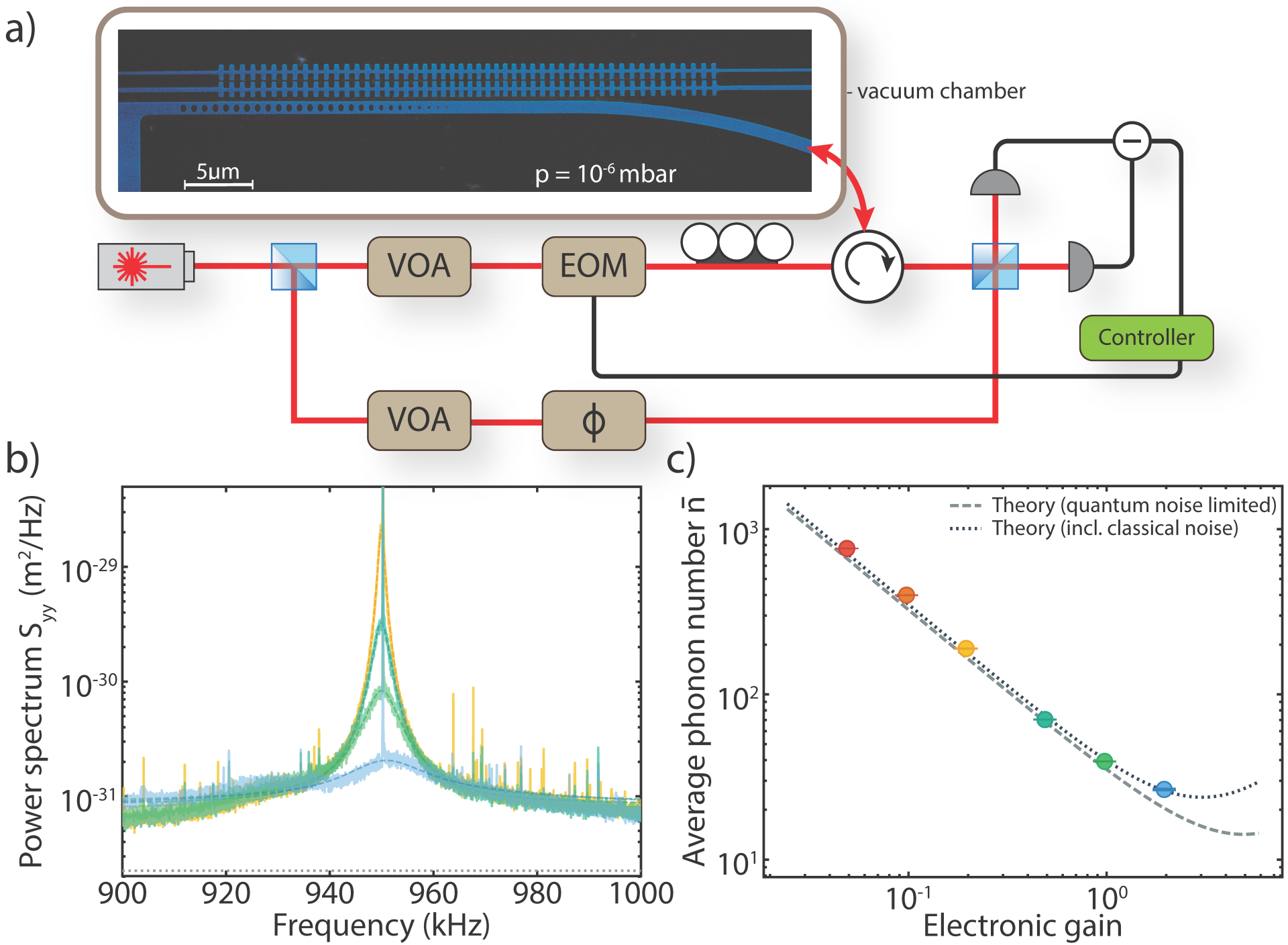}
	\caption{(a) Sketch of the feedback cooling setup. A laser is first tuned on cavity resonance and phase modulated to generate a calibration tone. It is then split into two arms, with both intensities being controlled through variable optical attenuators (VOA). The bottom path is the local  oscillator for the homodyne detection scheme, where the phase can be controlled using a fiber stretcher ($\phi$). The light in the upper (signal) arm is intensity modulated in an electro-optical modulator (EOM) and sent to the waveguide~\cite{Groeblacher2013a}, where it is then evanescently coupled into the optical cavity. At the end of the waveguide we pattern a photonic crystal mirror, which allows the light from the cavity to be reflected back into the fiber with a collection efficiency of 91\%. This light is then mixed with the local oscillator on a beamsplitter and measured in a home-built low-noise balanced photodetector with a quantum efficiency of 70\%, in order to perform the phase-sensitive measurement. The detected signal is electronically processed in an FPGA-based controller, which directly modulates the light in real-time through the EOM, and hence allows to cool the mechanical resonator. (b) Cooled mechanical spectra $S_{yy}$, with increased feedback gain from orange to blue and constant intracavity photon number $n\und{c}$. The dashed lines are fits to the spectra, while the gray dotted line indicates the quantum limited noise floor. (c) Average phonon number extracted from the spectra in (b), with corresponding color coding. The gray dashed line represents the theoretically predicted quantum-noise-limited phonon number, and the dark blue dotted line is the expected phonon number when taking the noise into account.}
	\label{fig:4}
\end{figure}

In this unresolved side band regime~\cite{Aspelmeyer2014}, an active feedback cooling scheme can be used to reduce the thermal energy of the mechanical oscillator~\cite{Kleckner2006b,Rossi2018}. In this scheme, unlike in the traditional cavity cooling approach~\cite{Chan2011}, the extremely large bandwidth of the optical cavity allows to retrieve information on the motion of the mechanical resonator at a high rate. In our experiment, we measure the position of the mechanics and process it in real-time, using the resulting signal to actively control the optical input power into the optomechanical cavity. The modulation of the intensity changes the radiation-pressure force, hence allowing to control and actively cool the mechanics itself. Figure~\ref{fig:4}a shows a sketch and description of our setup used to demonstrate such feedback cooling of our mechanical resonator. The measured signal containing the information on the position of the mechanical oscillator is sent to an FPGA controller (RedPitaya 125-14), with its output directly connected to an electro-optical intensity modulator just before the device. The FPGA control allows us to implement an almost arbitrarily complicated feedback filter. We apply a derivative filter with a $2^\mathrm{nd}$ order underdamped low-pass filter to cool the mechanical defect mode. The feedback phase at the resonance frequency is tuned to be $-\pi/2$. Due to a small delay in the system, applying this signal directly would heat other nearby mechanical modes and make the system unstable. We therefore cascade a series of notch filters to tune the phase response locally, which provides a weak cooling over the surrounding modes (cf.\ Figure~\ref{fig:3}). The total delay of the feedback loop is measured to be 0.49~$\mu$s.

Figure~\ref{fig:4}b shows the calibrated displacement power spectrum ($S_{yy}$) of the mechanical oscillator with different levels of cooling from a bath at room temperature. We keep the cavity photon number fixed at $n\und{c}=120$, while increasing the gain of the feedback filter to increase the amount of feedback. The mechanical peak amplitude reduces and broadens, corresponding to a cooling of the mode of interest. The curves are then fitted and we extract the displacement spectrum $S_{xx}$~\cite{Rossi2018, Krause2015}. This allows us to calculate the average phonon number $\bar{n}$, which is shown as a function of electronic gain in Figure~\ref{fig:4}c. The lowest occupation we obtain is $\bar{n}=26.6\pm0.7$, reduced from $6.5\times10^6$ at room temperature. We note that our measurements are not quantum-noise-limited in this experiment, resulting in a slightly increased phonon occupancy compared to the theoretically expected value. This additional noise floor results from the optical fiber touching the waveguide and introducing broadband mechanical modes. These mechanical waveguide modes shift the resonance frequency of the cavity weakly, resulting in an increase of the detection noise. At high gain, this noise is fed into the mechanics and limits the cooling efficiency. Unlike in the ideal case of a quantum-noise-limited measurement, increasing the input optical power does not reduce the classical noise and hence it does not lead to more efficient cooling. Using different types of coupling methods or re-designing the waveguide will allow us to reduce the classical noise further, allowing us to, in principle, cool to an occupation of $\bar{n}\und{min}\approx14$, with everything else left unchanged.

In order to get even closer to the groundstate in the continuous feedback cooling scheme, the measurement rate ($\Gamma\und{meas}=x\und{zpf}^2/S_{xx}^{\mathrm{imp}} = 4 \eta n\und{c} g_0^2  / \kappa$) has to be comparable or larger than the decoherence rates in the system, i.e.\ the thermal decoherence ($\Gamma\und{th} \approx \Gamma\und{m} n\und{th}$) and the back-action rate ($\Gamma\und{ba}=\Gamma\und{m}n\und{ba}$, where $n\und{ba}=n\und{c} C_0$)~\cite{Wilson2015}. Here, $\eta$ is the overall detection efficiency, which for our experiment is $\eta=0.50$, while $n\und{ba}=2.4\times10^4\ll n\und{th}$, hence making the thermal component the dominant decoherence channel. Excluding classical noise, we find $\Gamma\und{meas}/(\Gamma\und{th}/8)=0.015 \ll 1$~\cite{Wilson2015}, which is orders of magnitude larger than in previous similar experiments~\cite{Krause2015}.

Several approaches to increasing this ratio exist. For example, by redesigning the coupling waveguide to obtain a quantum-noise-limited measurement, the intracavity photon number can be raised further, and is eventually only bound by absorption heating. Increasing the optomechanical coupling rate can be achieved by improving the fabrication and reducing the gap size between the strings forming the optical cavity. A reduction of the gap to 100~nm yields $G\und{om}=2\pi \times 45$~GHz/nm, which is more than twice the current value. Another way is to further reduce the thermal decoherence rate, through device improvements. Our current design is not optimized to maximize the stress~\cite{Norte2016}, which would lead to more stored energy, increasing $Q\und{m}$. At the moment, the maximum simulated stress in the structure is 1.5~GPa, which is still far below the yield strength of SiN ($\sim$6~GPa). Higher stress can also be achieved through an overall longer beam, while at the same time allowing for more adiabatic chirping in the geometry, which would further reduce mechanical losses. Combining all of these approaches should allow to reach the quantum groundstate. In partiular, moderately increasing the mechanical quality factor to $Q\und{m}\approx1\times10^8$, would lead to an increase of $\Gamma\und{meas}/(\Gamma\und{th}/8)$ to 0.06 and the phonon number could be reduced to 6. Together with a reasonable reduction of the gap to 100~nm and a small increase of the cavity photon number to 200, a phonon occupation around 3 will be achievable. Further improvements in $Q\und{m}$ to $\gtrsim7\times10^8$~\cite{Tsaturyan2017,Ghadimi2018} will finally enable phonon numbers below unity starting from room temperature.

In summary, we have designed and fabricated a novel, fully integrated optomechanical system, featuring a fishbone-like photonic and phononic structure,  with a $Q\und{m}=2.73\times10^7$ of an in-plane mechanical mode combined with a large optomechanical coupling rate of $g_0=237$~kHz. We use this device to demonstrate active-feedback cooling close to the quantum groundstate of motion, starting from room temperature. By tuning the FPGA-based feedback filter, we stabilize spurious modes that strongly couple to the optics, allowing us to reach an effective mode temperature of 1.2~mK, corresponding to less than 27 phonons. Further improvements in the noise performance of our setup, together with enhancements of $Q\und{m}$ and optomechanical coupling, should allow for these structures to be cooled fully into their groundstate, which will enable mechanical quantum experiments at ambient temperatures. In addition, the simplicity in fabrication of our devices, consisting of a single SiN layer on chip only, combined with their fully integrated on-chip character, makes them ideal candidates for quantum sensing applications~\cite{Krause2015,Norte2018}.

\begin{acknowledgments}
We would like to thank Bas Hensen, Alex Krause, Igor Marinkovi\'{c}, Rob Stockill, and Andreas Wallucks for valuable discussions and also acknowledge assistance from the Kavli Nanolab Delft. This work is further supported by the Foundation for Fundamental Research on Matter (FOM) Projectruimte grants (15PR3210, 16PR1054), the European Research Council (ERC StG Strong-Q, 676842), the EMPIR programme co-financed by the Participating States and from the European Union's Horizon 2020 research and innovation programme, and by the Netherlands Organization for Scientific Research (NWO/OCW), as part of the Frontiers of Nanoscience program, as well as through a Vidi grant (680-47-541/994). This project is part of ATTRACT that has received funding from
the European Union's Horizon 2020 Research and Innovation Programme. J.G.\ gratefully acknowledges support through a Casimir PhD fellowship. 
\end{acknowledgments}

\setcounter{figure}{0}
\renewcommand{\thefigure}{S\arabic{figure}}
\setcounter{equation}{0}
\renewcommand{\theequation}{S\arabic{equation}}

\clearpage

\section{Supplementary Information}

\subsection{Device design}

For our design, we first set the geometry of the photonic crystal as the fishbone structure. We then fix the width of the narrow part of the photonic crystal to $w_{a}^{\mathrm{Pht}}=165$~nm, a width that can easily be fabricated with high yield. The photonic crystal consists of adiabatically chirped unit cells, with a defect region at the center and mirror regions at the ends. We then perform finite element simulations (FEM) of the optical properties of a unit cell of the mirror region and the defect, using COMSOL. The free parameters are the period of the unit cells $a^{\mathrm{Pht}}$ at the defect region and at the mirror region, as well as the width $\mathrm{w_1^{\mathrm{Pht}}}$ and the ratio between the length of the wide part $L_b^{\mathrm{Pht}}$ and the period $a^{\mathrm{Pht}}$. These parameters are tuned such that we obtain a bandgap around 1550~nm for TE-like modes for the mirror. We design the defect such that the lower-band crosses the center of this bandgap. We also set transition of the parameters to follow a Gaussian function, $p_n=p_0-(p_{\infty}-p_0)\mathrm{exp(-n^2/M^2)}$, where $p_n$ is the corresponding parameter of the n-th unit cell. Here $p_{\infty}$ is the parameter for the unit cell corresponding to the mirror region, while $p_0$ is the unit cell at the center. The parameter $M$ determines the adiabaticity of the transition, which we set to 9 in our case. With this initial design we then run an optimization algorithm~\cite{ChanPhD} maximizing $G\und{om}/\kappa$, where $G\und{om}$ is calculated using the moving boundary effect~\cite{Johnson2002}. For the structure we use in this work, the designed parameters are listed in Table~\ref{Tab:S1}.

\begin{table}[!htb]
	\centering
	\begin{tabularx}{1\columnwidth}{| X | X | X |}
		\hline
		Parameter            & $p_0$     & $p_\infty$  \\ \hline
		$a^{\mathrm{Pht}}$   & 590~nm               & 691~nm         \\ \hline
		$w_b^{\mathrm{Pht}}$         & 992~nm        & 992~nm         \\ \hline
		$L_\mathrm{b}^{\mathrm{Pht}} /a^{\mathrm{Pht}}$  & 0.461        & 0.399         \\ \hline
	\end{tabularx}
	\caption{Parameters of the fishbone photonic crystal for the center unit cell and unit cells at the mirror region. The definitions of the parameters can be found in Figure~\ref{fig:S_DevPara}.}
	\label{Tab:S1}
\end{table}

\begin{table}[!htb]
	\centering
	\begin{tabularx}{1\columnwidth}{| X | X | X | X | X |}
		\hline
		Parameter            & $p_0$ (fishbone)   & $p_\infty$ (fishbone) & $p_0$ (block) & $p_\infty$ (block) \\ \hline
		$a^{\mathrm{Mech}}$   & 283~$\mathrm{\mu}$m  & 91.6~$\mathrm{\mu}$m &  283~$\mathrm{\mu}$m & 91.6~$\mathrm{\mu}$m \\ \hline
		$w_a^{\mathrm{Mech}}$         & 170~nm   & 550~nm    & 170~nm   & 550~nm      \\ \hline
		$w_b^{\mathrm{Mech}}$         & 480~nm   & 746~nm    & 288~nm   & 746~nm      \\ \hline
		$L_b^{\mathrm{Mech}} /a^{\mathrm{Mech}}$  & 0.035        & 0.29 & -0.1 & 0.29         \\ \hline
	\end{tabularx}
	\caption{Parameters of the mechanical phononic structure, for which the definitions can be found in Figure~\ref{fig:S_DevPara}. The indices of the rectangular block phononic unit cells starts from 3, following the indices for the fishbone phononic structure. Thus a negative value appears for the block unit cells. The actual number of fins in the fishbone structure are rounded to the nearest integer.}
	\label{Tab:S2}
\end{table}

\begin{figure}[!htb]
	\includegraphics[width=1\columnwidth]{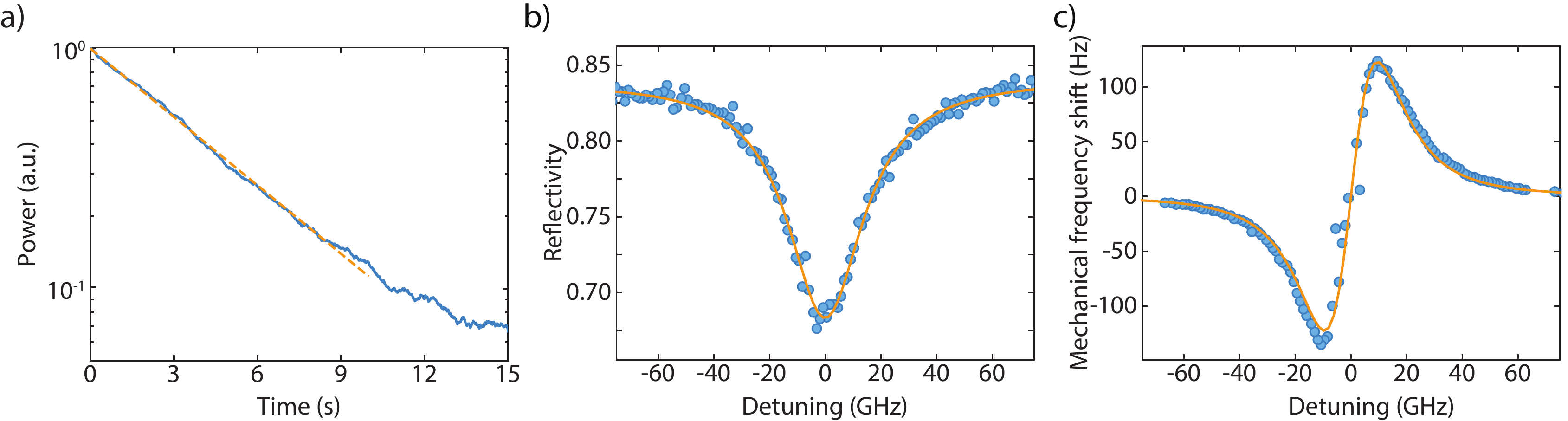}
	\caption{(a) Ringdown measurement of the mechanics with a decay time constant of 4.58~s, corresponding to a mechanical quality factor of $Q\und{m}=2.73\times10^7$ ($\omega\und{m}/2\pi=950.4$~kHz). (b) Normalized optical cavity resonance ($\lambda=1549.9$~nm). (c) Optical-spring effect measured by detuning the laser around $\omega\und{m}$. The fits to (b) and (c) yield a total cavity linewidth $\kappa/2\pi=33.0$~GHz and optomechanical coupling $G\und{om}/2\pi=21.6$~GHz/nm.}
	\label{fig:S1}
\end{figure}

\begin{figure*}
	\includegraphics[width=2.\columnwidth]{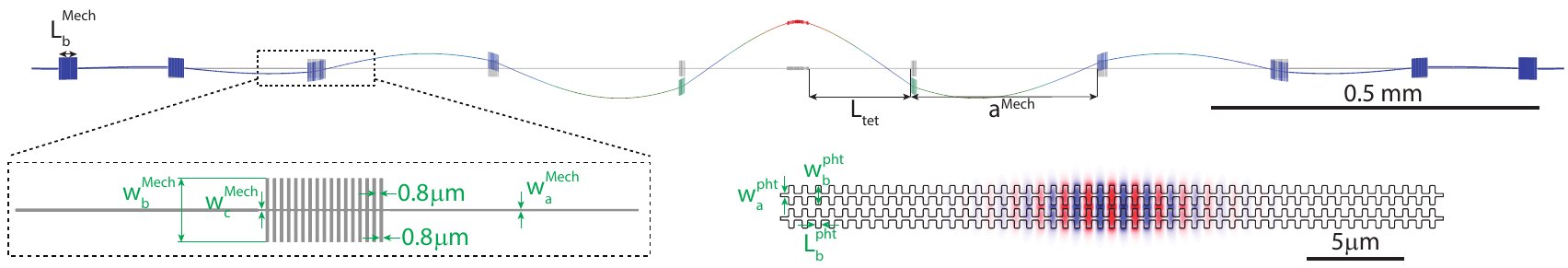}
	\caption{Definition of the parameters for the full structure.}
	\label{fig:S_DevPara}
\end{figure*}

The mechanical simulations are also performed using COMSOL, where the width of the initial string is again chosen to be 165~nm. We first simulate the band structure for the unit cell closest to the clamping region and a unit cell of the last phononic fishbone structure. In order to reduce the parameter space for the fishbone structure, both its vertical width and the distance between individual fins is fixed to 0.8~$\mu$m. Within the fishbone structure, the width of the string connecting the fins $w_c^\mathrm{Mech}$ is set to 165~nm for the first unit cell, and 250~nm for the mirror region. The parts of strings closest to the rectangular blocks are narrowed down to follow the same transition. Through the simulations, we find bandgaps for the block and fishbone structure, respectively. Finally, we run a full mechanical simulation of the whole assembled long string. For the device presented here, 3 unit cells on each side have the fishbone structure, while another 3 are rectangular blocks. As the stress differs between the individual unit cell and the complete simulations, due to the finite length of the structure, the actual bandgap of the device deviates slightly. We manually adjust the length of the string connecting to the photonic crystal $L_{\mathrm{tet}}$ in order to obtain the right mode. We then run another optimization, this time maximizing $f\und{m}\cdot Q\und{m}$, where $Q\und{m}$ is evaluated as the ratio between the stored elastic energy and the bending loss~\cite{GhadimiPhD}. We also set a bound on $f\und{m}$ in the optimization in order to limit the mechanical frequency range. The main parameters and the corresponding values are shown in Table~\ref{Tab:S2}. As before, the same Gaussian transition is used. parameters. Note that there is no adiabatic transition between the fishbone phononic structure and the rectangular blocks. Instead, we use two sets of parameters defining their geometries.

\begin{figure}
	\includegraphics[width=1.\columnwidth]{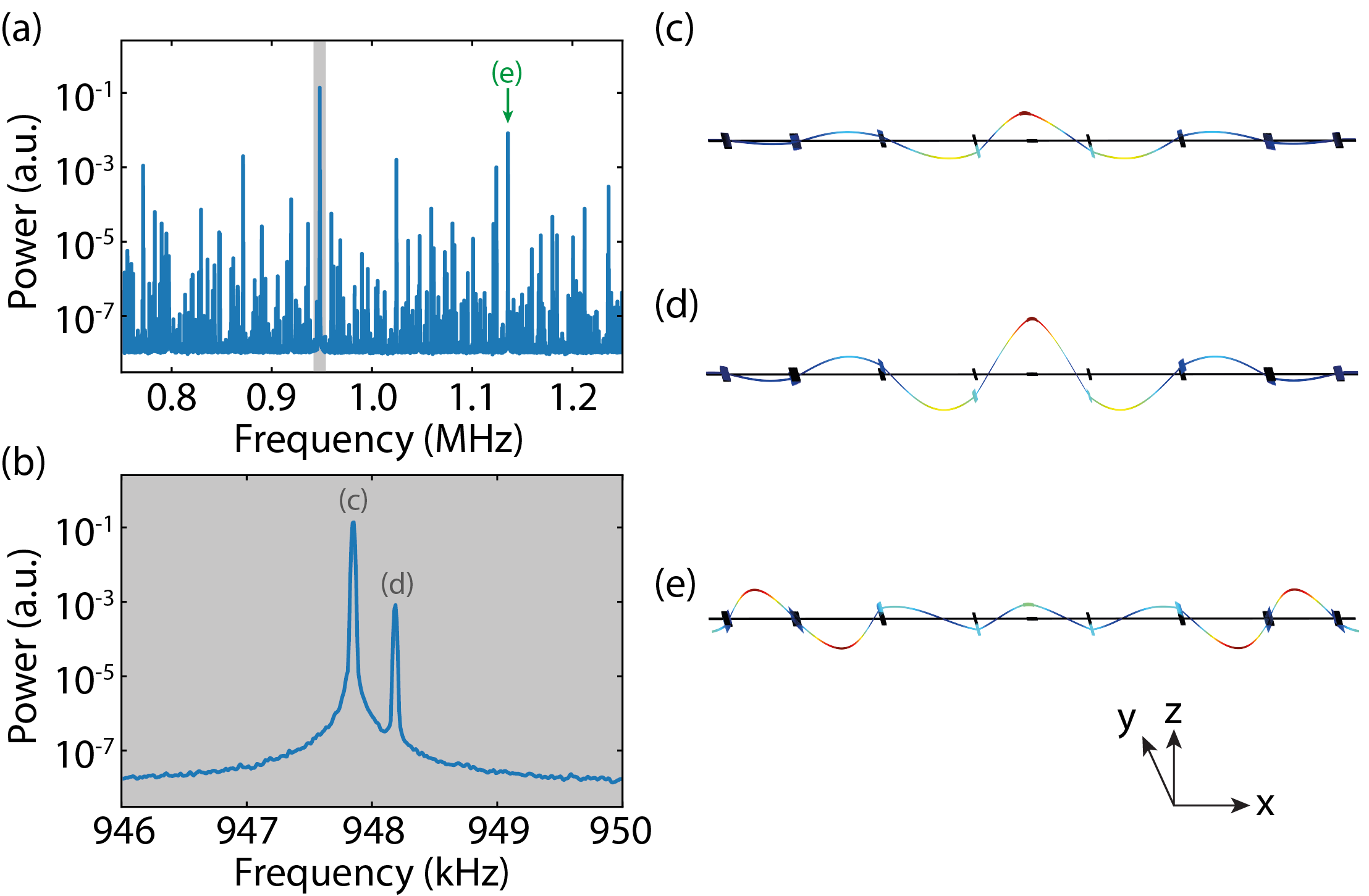}
	\caption{(a) Mechanical spectrum close to the high-Q mode. A zoom-in of the highlighted gray area is shown in (b). (c-e) Simulated mode shapes of the corresponding modes labeled in (a) and (b). The mode in (c) is the in-plane mode of interest, (e) a higher order mode, while (d) is out of plane.}\label{fig:S_MechModes}
\end{figure}

\subsection{Device fabrication \& characterization}

Our devices are fabricated from a 350~nm thick high-stress silicon nitride film deposited in an LPCVD furnace on silicon. The pattern is first generated using standard electron beam lithography and then transferred into the silicon nitride layer using a CHF$_3$ plasma etch. Next, we clean the chip in a piranha solution and use hydrofluoric acid to remove any oxidation. Finally, we undercut our structures using a fluorine-based dry release~\cite{Norte2018}. We would like to highlight the simplicity of the fabrication process, requiring a minimal amount of steps in order to make the suspended structure, completely  avoiding the need for complex multilayered processes~\cite{Ghadimi2017}.

In order to measure the mechanical and optical properties of our device, we perform a mechanical ringdown measurement, as well as a wavelength sweep of the optical resonance (see Figure~\ref{fig:S1}a,b), respectively. We determine the optomechanical coupling rate by measuring the optical spring effect of the mechanical mode as a function of laser detuning (cf.\ Figure~\ref{fig:S1}c).

The mechanical spectrum close to our mode of interest is shown in Figure~\ref{fig:S_MechModes}a, from which we identify and verify that the mechanical mode we measure is the correct in-plane mechanical mode we design. A zoom-in of the spectrum around 948~kHz (gray region) is plotted in Figure~\ref{fig:S_MechModes}b. Due to the single-layer geometry, the in-plane mode (left) has a large optomechanical coupling, while the coupling of the out-of-plane mode (right) is weak. The out-of-plane mode has a slightly higher frequency, which agrees with our simulations (see c and d). We further confirm the mode by comparing it with a higher order in-plane mode (green arrow, Figure~\ref{fig:S_MechModes}e), whose frequency is 19~kHz higher, which is also in excellent agreement with simulations (20~kHz higher).

\vspace*{5.cm}

\begin{figure}
	\includegraphics[width=1.\columnwidth]{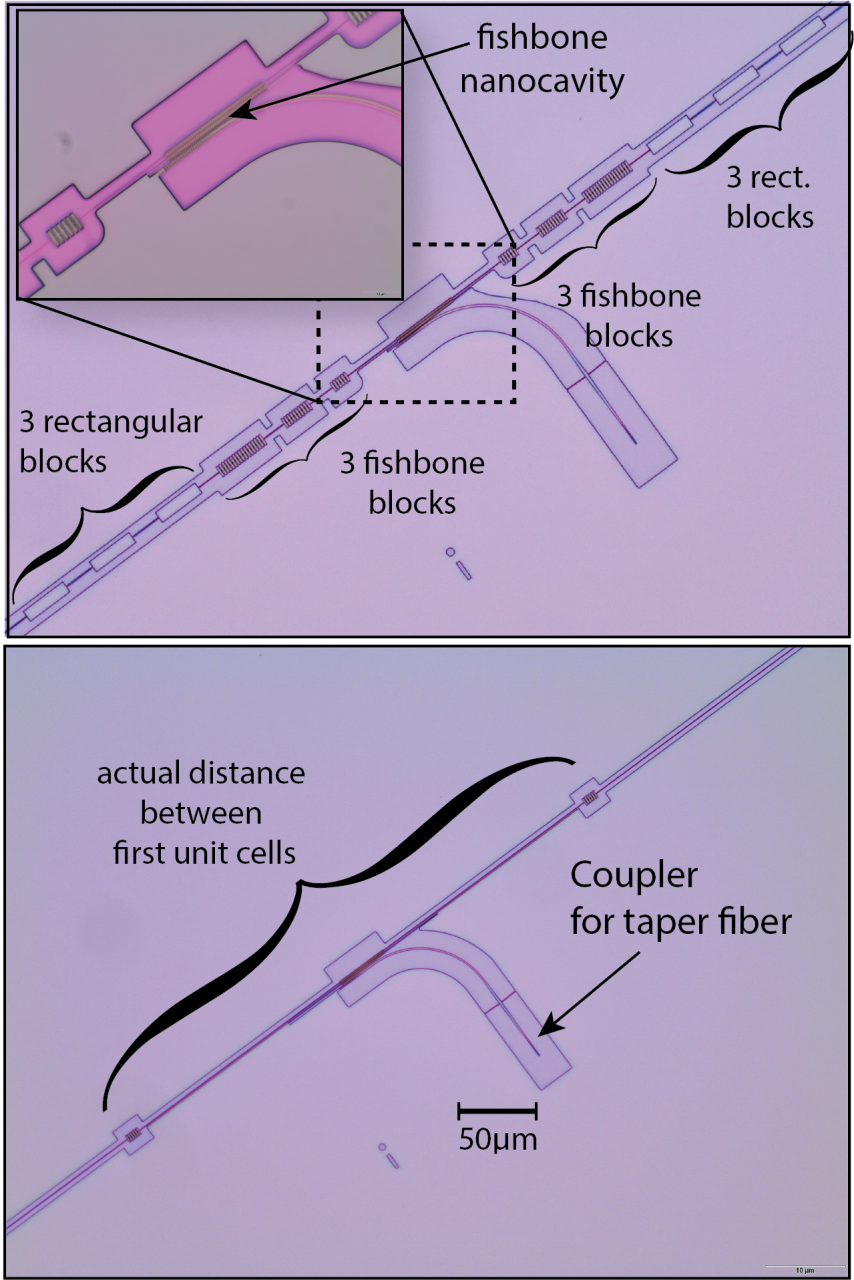}
	\caption{(top) Microscope image of the complete device design. The tethers of the structure shown here are shortened by about a factor of 10 compared to the device used in the main text in order to fit into one picture. The long beam consists of the fishbone nanocavity structure in the center, symmetrically connected on both sides by 3 fishbone phononic shield blocks, followed by 3 rectangular phononic shield blocks and then clamped to the chip. The inset shows a zoom-in of the fishbone nanocavity. (bottom) Shown is an image of the device with the actual dimensions used in our experiments, with only the first fishbone phononic shield blocks visible. Both images show the waveguide used to couple light from the tapered optical fiber into the photonic crystal cavity.} \label{fig:S2}
\end{figure}

\end{document}